\documentclass[twocolumn]{autart}
\usepackage[noadjust]{cite}
\usepackage{amsmath,amsfonts,amssymb,graphicx}
\usepackage{algorithm}
\usepackage{pdfsync,keyval}
\usepackage{booktabs}
\usepackage{algpseudocode}
\usepackage{graphics}
\usepackage{epsfig}
\usepackage{multirow}
\usepackage{subfigure}
\usepackage{tikz}
\usepackage{bm}
\usepackage[english]{babel}

\newtheorem{theorem}{Theorem}[section]
\newtheorem{lemma}[theorem]{Lemma}
\newtheorem{corollary}[theorem]{Corollary}

\newtheorem{proposition}[theorem]{Proposition}
\newtheorem{remark}[theorem]{Remark}

\newtheorem{assumption}[theorem]{Assumption}

\newcommand{\hfs}{\hfill\ensuremath{\square}}

\begin{document}

\begin{frontmatter}

\title{Secure Analysis of Dynamic Networks under Pinning Attacks against Synchronization}

\author[NEU]{Yuzhe Li}\ead{yuzheli@mail.neu.edu.cn},
\author[BIT]{Dawei Shi}\ead{dawei.shi@outlook.com},
\author[UA]{Tongwen Chen}\ead{tchen@ualberta.ca}

\thanks{This manuscript is the initial submitted version of a paper with the same title to be published in Automatica, which has been revised and condensed. }

\address[NEU]{State Key Laboratory of Synthetical Automation for Process Industries, Northeastern University, Shenyang, 110004, China.}
\address[BIT]{State Key Laboratory of Intelligent Control and Decision of Complex Systems, School of Automation, Beijing Institute of Technology, Beijing, 100081, China.}
\address[UA]{Department of Electrical and Computer Engineering, University of Alberta, Edmonton, AB, T6G 1H9, Canada .}

%
%
%
\maketitle

\begin{abstract}
In this paper, we first consider a pinning node selection and control gain co-design problem for complex networks. A necessary and sufficient condition for the synchronization of the pinning controlled networks at a homogeneous state is provided. A quantitative model is built to describe the pinning costs and to formulate the  pinning node selection and control gain design problem for different scenarios into the corresponding optimization problems. Algorithms to solve these problems efficiently are presented. Based on the developed results, we take the existence of a malicious attacker into consideration and a resource allocation model for the defender and the malicious attacker is described. We set up a leader-follower Stackelberg game framework to study the behaviour of both sides and the equilibrium of this security game is investigated. Numerical examples and simulations are presented to demonstrate the main results.
\end{abstract}


\end{frontmatter}

\section{Introduction}

Complex networks refer to a general type of networks consisting of huge numbers of interconnected nodes, where each node can be regarded as an individual dynamical system coupled with other nodes. Over the past decades, intensive research efforts  have revealed that complex networks provide suitable mathematical models to describe certain real-world systems, e.g., social networks, biological networks, power grids, and neural networks \cite{rubinov2010complex,sayed2014adaptive,zhang2006complex}. Take the swarm behaviour (e.g.,  bird flocks, fish schools, and insect swarms) in biological networks as an example; in such a self-organized phenomenon, each participant in the group makes decisions and takes actions based on the information exchange among its near neighbors, leading to a collective behaviour of movement exhibited by the overall organization \cite{camazine2003self}. Naturally, understanding the mechanisms behind the swarm behaviour has become an interesting topic in the related areas \cite{reynolds1987flocks,ballerini2008interaction}. In fact,  the synchronization and collective control problems are of essential importance not only in the biological areas, but also in the investigations of all kinds of complex networks \cite{watts1998collective,lu2005time,motter2005enhancing,chen2009reaching,wang2010control}.

In general, due to the consideration of a large number of nodes, it is difficult to control all the individual dynamics in a complex network at the same time to accomplish certain tasks, which may induce huge implementation costs. Instead, the technique of the so-called ``pinning control" has emerged and been investigated in the literature in recent years \cite{grigoriev1997pinning,gang1994controlling,wang2002pinning,song2012pinning,chen2009reaching,xiang2007pinning,porfiri2008criteria}. To be specific, the pinning control technique aims to lead the entire network to an objective state by applying local feedback controllers to a small portion of network nodes. The criteria and conditions for pinning-controllability for different types of networks are investigated thoroughly in \cite{wang2002pinning} (scale-free networks), \cite{chen2009reaching} (multi-agent systems), \cite{lu2005time} (time-varying networks), \cite{xiang2007pinning} (networks with general coupling topologies), and \cite{porfiri2008criteria} (global asymptotic stability). Specifically, the results in \cite{porfiri2008criteria} provide a sufficient condition for global-pinning controllability, which is determined by the network topology, the location of pinning nodes, the coupling strength, and the feedback gain. However, the optimal selection of the pinning nodes among the networks remains an open problem. In \cite{wang2002pinning}, the authors proposed two pinning schemes for scale-free networks: one is a randomly scheme and the other one is to choose the most highly-connected nodes, which have a better performance in terms of control costs. Similar schemes were widely adopted in the literature, e.g., \cite{chen2009reaching,xiang2007pinning}. By using a M-matrix approach, the results in \cite{song2012pinning} suggested that the interaction diagraph could be partitioned into a minimum number of components and the pinning control could be achieved if and only if the root node of each component was pinned. Typically, due to the various locations and connections of nodes, the costs for applying the control on them are different. In addition, given the same selection of pinning nodes, different control gains can also lead to different control costs. Therefore, the selection of pinning nodes and the design of the corresponding control gains can be optimized to achieve better overall network performance.

On the other hand, although most systems modeled by complex networks are benefited from the advance of modern technologies in terms of the information exchange, the interactive characteristics of both the cyber and the physical parts of complex networks bring new challenges to maintain the secure operation of complex networks \cite{motter2002cascade,callaway2000network,noel2005understanding}.  As shown in \cite{wang2002pinning,watts1998collective}, certain types of complex networks, e.g., the so-called ``small-world'' networks,  are robust to the removal of nodes. However, for most complex networks, the random failure or adversarial compromise  of nodes can lead to severe damage to the networks operation and functionality \cite{motter2002cascade}. To defend the malicious attacks on the nodes of complex networks, one straightforward strategy is to secure the nodes in the networks. Take the so-called ``smart grid" as an example: As the future electricity network consists of a huge number of ``prosumers'' (consumers and producers of energy), its smooth operation is closely related to the national economy and security \cite{pagani2014power} and can be secured by installing monitoring cameras or alarm systems, and upgrading the sensor, communication and control infrastructures \cite{deng2015defending}. Due to the large number of nodes in the networks and a limited defensive budget, the defender of the networks needs to balance the resource allocation among the nodes. The malicious attacker is also confronted with a similar resource constraint and may compromise the nodes selectively, namely, via ``pinning attacks''. To model the situation where multiple agents make decisions interactively under constraints, game-theoretic methods are widely used in the literature \cite{poncela2009evolutionary,alpcan2010network}. Under the game framework, how the defender of the complex networks allocates the limited resources among nodes and how the malicious attacker designs the strategy of pinning attacks are interesting and important problems to investigate.

Motivated by the existing results in the literature, in this paper, we first consider a pinning node selection and control gain design problem for complex networks. We provide a necessary and sufficient
condition for a pinning controlled network to synchronize at a homogeneous state. A quantitative model is built to describe the pinning costs and formulate the  pinning node selection and control gain design problem in different scenarios as corresponding optimization problems. Algorithms to solve these problems efficiently are also presented. Based on the developed results, we then take the malicious attacker into consideration.  A resource allocation model for the complex network defender and the malicious attacker is described. Different from the analysis in terms of the so-called ``Nash equilibrium", where both sides act  {\it simultaneously}, it is more reasonable that the decisions are made {\it sequentially}: typically the defender first decides its defense resource allocation, and the attacker observes and chooses nodes to launch pinning attacks. Therefore, we set up a leader-follower framework (namely, the so-called ``Stackelberg game" framework \cite{korzhyk2011stackelberg}) to study the behaviour of both sides and the equilibrium of this security game is investigated. The main contributions of the current work are summarized as follows:
\begin{enumerate}
\item A pinning node selection and control gain co-design problem for complex networks in the presence of a malicious attacker is studied. To the best of our knowledge, the problem formulation and results are new.
\item The interested pinning problems for different scenarios are formulated into corresponding quantitative optimization problems. The solution to the case of free selection of nodes are provided in Theorem~\ref{thm:convex1}; the solution to a special case when identical control
gains are adopted is provided in Theorem~\ref{thm:BLP1}, while a  Branch-and-Bound algorithm (Algorithm~\ref{algm:BB}) to solve the problem efficiently is also presented; the solution to the case of constrained number of pinning nodes is provided in Theorem~\ref{thm:convex_cardinality}.
\item A Stackelberg security game between the defender of the complex network and an adversary launching pinning attacks on the nodes is studied and the equilibrium of the game is investigated. Specifically, the solutions to the Stackelberg security game for both the defender and the attacker are provided in Theorem~\ref{thm:Solution_stackelberg};  the solutions for the cases of fixed budget constraints for both the defender and the attacker are also studied in Proposition~\ref{prop:Pi_d} and Proposition~\ref{prop:Pi_a}, respectively.
\end{enumerate}

The remainder of the paper is organized as follows. Section II presents the preliminaries on the pinning control of complex networks. The pinning node selection and control gain co-design problems are investigated in Section III. Section IV provides the analysis of the proposed Stackelberg game. Numerical examples and simulations are demonstrated in Section V. Section VI concludes the work with several remarks.

\textit{Notations}: $\mathbb{Z}$ and $\mathbb{N}^+$ (or $\mathbb{N}$) denote the set of all integers and positive integers (or non-negative integers), respectively. $\mathbb{R}$ is the set of real numbers and $\mathbb{R}^+$ is the set of positive real numbers. $\mathbb{R}^{n}$ is the $n$-dimensional Euclidean space. When $X$ is a positive semi-definite matrix (or positive definite matrix), we write $X \geqslant 0$ (or $X > 0$). We write $X\geqslant Y$ (or $X > Y$) if $X - Y$ is a positive semi-definite matrix (or positive definite matrix). $\mathrm{Tr}\{\cdot\}$ is the trace of a square matrix. The superscript~$'$ stands for transposition. $[a_{ij}]$ denotes a matrix with $a_{ij}$ as the element in the $i$-th row and the $j$-th column. For functions $f, f_1, f_2$ with appropriate domains, $f_1\circ f_2(x)$ stands for the function composition $f_1\big (f_2(x)\big)$, and $f^{n}(x) \triangleq f\big(f^{n-1}(x)\big)$, where $n\in\mathbb{N}$ and with $f^{0}(x) \triangleq x$. $\bm 1\triangleq[1,1,...,1]'$ and $\bm I$ is the identity matrix, with proper dimensions when no ambiguity arises. $\delta_{ij}$ is the Dirac delta function, i.e., $\delta_{ij}$ equals to $1$ when $i=j$ and $0$ otherwise. The notation $\mathbb{P}[\cdot]$ refers to probability and $\mathbb{E}[\cdot]$ to expectation. $X_{i:j}$ and $\{X_l\}_{l=i}^j$ represent the same set $\{X_i,X_{i+1},...,X_{j}\}$. The cardinality of a set $\mathbb A$ (or a vector $x$) is denoted  as $|\mathbb A|$ (or $\textbf{card}(x)$).

\section{Preliminaries}

\subsection{Complex Dynamical Networks}

We consider a general complex network with $N$ identical coupled nodes, and each node can be represented as an $n$-dimensional dynamical system with state equations as
\begin{equation} \label{eqn:system_model}
   \dot{\bm x}_i=\bm f(\bm x_i)+c\sum_{j=1}^Na_{ij}\bm g(\bm x_j),~~i\in\mathcal V\triangleq\{1,2,...,N\},
\end{equation}
where $\bm x_i=[x_{i,1},x_{i,2},...x_{i,n}]'\in \mathbb R^{n}$ is the state vector of the $i$-th node; the continuously differentiable functions $\bm f=[f_1,f_2,...,f_{n}]': \mathbb R^{n} \mapsto \mathbb R^{n}$ and $\bm g=[g_1,g_2,...,g_{n}]': \mathbb R^{n} \mapsto \mathbb R^{n}$ represent the local dynamics of each nodes and the dynamics of inner-coupling from other nodes, respectively; the constant scalar $c>0$ describes the coupling strength of each node; the binary variable $a_{ij}$ $(i\neq j)$ denotes the coupling relation between different nodes: if node $i$ and node $j$ are connected to each other, then $a_{ij}=a_{ij}=1$; otherwise, $a_{ij}=a_{ij}=0$. Based on the classic graph theory, we denote the degree of node $i$, namely, the number of connections to node $i$, as $k_i$:
\begin{equation}
  k_i\triangleq\sum_{j=1,j\neq i}^Na_{ij},~~i\in\mathcal V,
\end{equation}
and
\begin{equation}
  a_{ii}\triangleq-k_i,~~i\in\mathcal V.
\end{equation}
As a result, we can further define the coupling matrix (Laplacian matrix) $\mathcal A\triangleq [a_{ij}]\in \mathbb R^{N\times N}$ to describe the coupling structure of the network. Assume that  the network is connected in the sense that there are no isolate clusters, and we have the following results.

\begin{lemma}[\cite{wang2002pinning}] \label{lem:propoerty_matrix_A}
   The coupling matrix $\mathcal A$ has the following properties:
  \begin{enumerate}
    \item $\mathcal A$ is a symmetric irreducible matrix with $n$ real eigenvalues;
    \item $\mathcal A$ has an eigenvalue $0$ with multiplicity 1 and all the other eigenvalues are strictly negative.
  \end{enumerate}
\end{lemma}

\begin{remark}
   Similar models as the one in \eqref{eqn:system_model} are widely adopted in the literature \cite{wang2002pinning,wang2010control,xiang2007pinning,chen2009reaching}. Different selections of the functions $\bm f$ and $\bm g$ in \eqref{eqn:system_model} can describe a variety of network behaviour and coupling schemes, e.g., fixed points, periodic orbits and chaotic states \cite{xiang2007pinning}. In particular, when the nodes in the network are linearly coupled, i.e., $\bm g(\bm x)=\bm x$, the dynamics in \eqref{eqn:system_model} becomes the one considered in \cite{wang2002pinning,wang2010control}. \hfs
\end{remark}

\subsection{Pinning Control of Networks}

The goal of pinning control is to drive the network in \eqref{eqn:system_model} to a homogeneous state $\bar {\bm x}$ via pinning local feedback controllers to a small fraction $\delta$ $(0<\delta<1)$ of the nodes, i.e.,
\begin{equation}
  \bm x_i=\bar {\bm x},~~i\in\mathcal V,
\end{equation}
and
\begin{equation}
  \bm f(\bar {\bm x})=0.
\end{equation}

Assume that there are $l$ nodes to be pinned, where $l$ is the smaller but nearest integer to the real number $\delta N$. Let $\mathcal V_{\text{pin}}\subseteq\mathcal V$ denote the set of pinning nodes and we can express the pinning controlled network as:
\begin{equation} \label{eqn:system_model_pinning}
  \dot{\bm x}_i=\bm f(\bm x_i)+c\sum_{j=1}^Na_{ij}\bm g(\bm x_j)+d_i\bm u_i,~~i\in\mathcal V,
\end{equation}
where the local feedback controller $\bm u_i$ (with control gain $c_i>0$) is given by:
\begin{equation} \label{eqn:controller_i}
  \bm u_i=-c_i\big[\bm g(\bm x_i)-\bm g(\bar {\bm x})\big],~~i\in\mathcal V,
\end{equation}
and the binary variable $d_i$ ($\bm d\triangleq [d_1,d_2,...,d_N]'$) indicates the selection of pinning nodes:
\begin{equation}
  d_i=\begin{cases}
   1, &i\in\mathcal V_{\text{pin}},\\
   0, &i\in\mathcal V\setminus \mathcal V_{\text{pin}}.
     \end{cases}
\end{equation}

\begin{remark}
  Notice that the local controller model in \eqref{eqn:controller_i} extends the ones in \cite{wang2002pinning,wang2010control,xiang2007pinning}, which adopt an identical control gain for all the nodes. \hfs
\end{remark}

To facilitate the later discussion, we define two matrices:
 \begin{equation} \label{def:mathcal_C}
\mathcal C\triangleq{\rm diag}\Big\{\frac{c_1 d_1}{c},\frac{c_2 d_2}{c},...,\frac{c_N d_N}{c}\Big\}\in\mathbb R^{N\times N},
 \end{equation}
 and
 \begin{equation}\label{def:mathcal_B}
 \mathcal B \triangleq \mathcal {A-C}\in\mathbb R^{N\times N}.
 \end{equation}
We further denote:
\begin{equation*}
    J[\bm f(\bar {\bm x})]\triangleq[\partial f_i(x)/\partial x_j]\in\mathbb R^{n\times n},
  \end{equation*}
and
\begin{equation*}
  J[\bm g(\bar {\bm x})]\triangleq[\partial g_i(x)/\partial x_j]\in\mathbb R^{n\times n},
\end{equation*}
as the Jacobian matrices of functions $\bm f$ and $\bm g$ at the state $\bar {\bm x}$, respectively.

The following theorem provides a necessary and sufficient condition for the pinning controlled network in \eqref{eqn:system_model_pinning} to synchronize at the homogeneous state $\bar {\bm x}$.

\begin{theorem} \label{thm:pinning_controllability}
  The pinning controlled network in \eqref{eqn:system_model_pinning} synchronizes at the homogeneous state $\bar {\bm x}$ if and only if all the $N$ matrices $J[\bm f(\bar {\bm x})]+c{\mu_{i}}J[\bm g(\bar {\bm x})]$, $i\in\mathcal V$, are stable in the sense that their eigenvalues all have negative real parts, where $\mu_{i}$, $i\in\mathcal V$, are the $N$ eigenvalues of the matrix $\mathcal B$.

\begin{pf}
To investigate synchronization of the pinning controlled network in \eqref{eqn:system_model_pinning}, first we define the error vector for node $i$ as:
\begin{equation}
  \bm e_i=\bm x_i-\bar {\bm x}.
\end{equation}
Clearly, to reach the homogeneous state $\bar {\bm x}$, we require that:
\begin{equation} \label{eqn:e_i_to_0}
  \lim_{t\to\infty}\bm e_i\to 0,~~\forall i\in\mathcal V.
\end{equation}

Similar to the procedures as in \cite{wang2002pinning,xiang2007pinning}, we can linearize \eqref{eqn:system_model_pinning} at the state $\bar {\bm x}$ and re-write the network dynamics in terms of $\bm e_i$. To be more specific, by using the properties of $\sum_{j=1}^Na_{ij}=0$ and $\bm f(\bar {\bm x})=0$, we have:
\begin{equation} \label{eqn:system_model_pinning_error}
  \dot{\bm e}_i=J[\bm f(\bar {\bm x})]\bm e_i+c\sum_{j=1}^N\Big(a_{ij}-\delta_{ij}\frac{c_i d_i}{c}\Big)J[\bm g(\bar {\bm x})]\bm e_j,~~i\in\mathcal V.
\end{equation}

Define
\begin{align*}
  \mathcal E&=[\bm e_1,\bm e_2,...,\bm e_N]\in\mathbb R^{n\times N},
\end{align*}
then we can present \eqref{eqn:system_model_pinning_error} in a more compact expression as:
\begin{equation} \label{eqn:system_model_pinning_error_compact}
  \dot{\mathcal E}=J[\bm f(\bar {\bm x})]\mathcal E+cJ[\bm g(\bar {\bm x})]\mathcal E\mathcal B.
\end{equation}

Notice that the matrix $\mathcal B$ is symmetric, thus $\mathcal B$ can be factorized in its eigendecomposition form  as:
\begin{equation} \label{eqn:decoposition_B}
  \mathcal {B} =\mathbf{\Phi}\mathbf{\Lambda}\mathbf{\Phi}^{-1},
\end{equation}
where $\mathbf{\Lambda}={\rm diag}\{\mu_1,\mu_2,...,\mu_N\}\in\mathbb R^{N\times N}$ is a matrix whose diagonal elements are the eigenvalues of matrix  $\mathcal B$ (without loss of generality, we assume that $\{\mu_i\}_{i=1}^n$ are arranged in non-decreasing order) and $\mathbf{\Phi}=[\bm \phi_1,\bm \phi_2,...,\bm\phi_N]\in\mathbb R^{N\times N}$ is an eigenvector basis satisfying $\mathbf{\Phi}'\mathbf{\Phi}=\bm I$, whose $i$-th column is the corresponding eigenvector of $\mu_i$.

Consider a transformation of $ {\mathcal E}$ on the basis $\mathbf{\Phi}$ and define $\bm \xi_i= {\mathcal E}\phi_i\in\mathbb R^{n}$. Thus we have
\begin{equation} \label{eqn:E_XI}
  {\mathcal E}=\bm \Xi\mathbf{\Phi}^{-1},
\end{equation}
where
\begin{equation*}
  \bm \Xi \triangleq [\bm \xi_1,\bm \xi_2,...,\bm \xi_N]\in\mathbb R^{n\times N}.
\end{equation*}
By substituting \eqref{eqn:decoposition_B} and \eqref{eqn:E_XI} into \eqref{eqn:system_model_pinning_error_compact}, we obtain that
\begin{equation*}
  \dot{\bm \Xi}\mathbf{\Phi}^{-1}=J[\bm f(\bar {\bm x})]\bm \Xi\mathbf{\Phi}^{-1}+cJ[\bm g(\bar {\bm x})]\bm \Xi\mathbf{\Phi}^{-1}\mathbf{\Phi}\mathbf{\Lambda}\mathbf{\Phi}^{-1},
\end{equation*}
i.e.,
\begin{equation} \label{eqn:system_model_pinning_error_compact_XI}
  \dot{\bm \Xi}=J[\bm f(\bar {\bm x})]\bm \Xi+cJ[\bm g(\bar {\bm x})]\bm \Xi\mathbf{\Lambda},
\end{equation}
which can be further re-written as:
\begin{equation} \label{eqn:system_model_pinning_error_compact_XI_i}
  \dot{\bm \xi_i}=\Big(J[\bm f(\bar {\bm x})]+c{\mu_i}J[\bm g(\bar {\bm x})]\Big)\bm \xi_i,~~i\in\mathcal V.
\end{equation}

It is straightforward that:
\begin{equation} \label{eqn:xi_i_to_0}
  \lim_{t\to\infty}\bm e_i\to 0 ~\Longleftrightarrow~ \lim_{t\to\infty}\bm \xi_i\to 0 ,~~\forall i\in\mathcal V.
\end{equation}

Clearly, to achieve the condition that $\lim_{t\to\infty}\bm \xi_i\to 0$, $\forall i\in\mathcal V$, it requires all the $N$ matrices $J[\bm f(\bar {\bm x})]+c{\mu_{i}}J[\bm g(\bar {\bm x})]$, $i\in\mathcal V$, to be stable, in the sense that every eigenvalue has a negative real part. Therefore, the synchronization condition of the pinning controlled network in \eqref{eqn:system_model_pinning} is equivalent to the stability of all  $N$ matrices $J[\bm f(\bar {\bm x})]+c{\mu_{i}}J[\bm g(\bar {\bm x})]$, $i\in\mathcal V$, which completes the proof.
\end{pf}
\end{theorem}

\section{Pinning Node Selection and Control Gain Co-Design}

In the previous section, we mainly focused on  synchronization condition of the pinning controlled network in \eqref{eqn:system_model_pinning} by taking the pinning scheme $\mathcal C$ as pre-determined. However, in many real world applications, it is also important to choose which nodes to be pinned. In this section, we look into this pinning node selection and control gain co-design problem, which serves as the basis for investigating the secure design problem in the presence of a malicious attacker in the next section.

\subsection{Preliminaries}
Before proceeding to the main results, let $\{\lambda_i(\cdot)\}_{i=1}^l$ denote the eigenvalues of the corresponding matrix arranged in non-decreasing order and we present the following supporting lemmas.

\begin{lemma}[\cite{horn2012matrix}] \label{lem:propoerty_eigen_M_N}
   Let two matrices $\mathcal M, \mathcal N \in\mathbb R^{l\times l}$ be symmetric, then we have:
     \begin{enumerate}
    \item $\lambda_i(\mathcal M+c\bm I)=\lambda_i(\mathcal M)+c$,
    \item $\lambda_i(\mathcal M+\mathcal N)\leqslant \lambda_{i+j}(\mathcal M)+\lambda_{n-j}(\mathcal N)$, $j=0,1,...,l-i$,
    \item $\lambda_i(\mathcal M+\mathcal N)\geqslant\lambda_{i-j+1}(\mathcal M)+\lambda_{j}(\mathcal N)$, $j=1,2,...,i$.
  \end{enumerate}
\end{lemma}

\begin{lemma}
The largest eigenvalue of the matrix $\mathcal B$ is non-positive:
\begin{equation*}
  \mu_N\leqslant0,
\end{equation*}
and all its other eigenvalues  are  strictly negative. Particularly, when $\mathcal V_{\text{pin}}=\mathcal V$, i.e., $d_i=1$, $\forall i\in\mathcal V$, $\mu_N$ is also strictly negative.
\begin{pf}
  From Lemma \ref{lem:propoerty_matrix_A}, we can conclude that:
  \begin{equation*}
    \lambda_N(\mathcal A)=0,~~\lambda_i(\mathcal A)<0,~i=1,2,3,...,N-1.
  \end{equation*}

Since $c_i>0$ and $d_i\in\{0,1\}$, $\forall i\in\mathcal V$, all the eigenvalues of the matrix $\mathcal C$ are non-negative, thus we have:
\begin{equation*}
  \lambda_N(-\mathcal C)=-\lambda_1(\mathcal C)\leqslant0.
\end{equation*}

Therefore, from Lemma \ref{lem:propoerty_eigen_M_N}, we have:
  \begin{align*}
    \mu_i&=\lambda_i(\mathcal B)\\
    &=\lambda_i[\mathcal A+(-\mathcal C)]\\
    &\leqslant \lambda_{i}(\mathcal A)+\lambda_{N}(-\mathcal C)<0, i\in\mathcal V\setminus \{N\},
  \end{align*}
and
  \begin{equation*}
    \mu_N\leqslant \lambda_{N}(\mathcal A)+\lambda_{N}(-\mathcal C)\leqslant0.
  \end{equation*}

When $\mathcal V_{\text{pin}}=\mathcal V$, we have $d_i=1$, $\forall i\in\mathcal V$. Thus there exists a certain $j\in\mathcal V_{\text{pin}}$, such that:
  \begin{equation*}
   \lambda_N(-\mathcal C)=-\lambda_1(\mathcal C)=-\frac{c_j d_j}{c}<0,
  \end{equation*}
which makes $\mu_N$ strictly negative and completes the proof.
\end{pf}
\end{lemma}

\begin{remark} \label{rmk:factors}
  Theorem~\ref{thm:pinning_controllability} summarizes the conditions to ensure synchronization of the pinning controlled network, which depends on various factors, including the network coupling structure $\mathcal A$, the coupling strength $c$, the nodes selection $d_i$, the  control gain $c_i$, the local dynamics $\bm f$ and the inner-coupling dynamics $\bm g$. However, even when all the nodes are connected in a preferable structure and can be controlled with arbitrary control gains, the pinning controlled network in \eqref{eqn:system_model_pinning} may still not synchronize at a homogeneous state. To illustrate this, consider a special case when $J[\bm f(\bar {\bm x})]=2\bm I$ and $J[\bm g(\bar {\bm x})]=-\bm I$. Since $\mu_N\leqslant 0$, all the matrices $J[\bm f(\bar {\bm x})]+c{\mu_{i}}J[\bm g(\bar {\bm x})]$, $i\in\mathcal V$, are unstable. As a result, the pinning controlled network may not synchronize  at a homogeneous  state. \hfs
\end{remark}

Based on the arguments in Remark~\ref{rmk:factors}, to focus on the pinning node selection problem, we develop our subsequent results  based on the following assumptions, although we will show later that the assumption on $\bm g$ can be relaxed.
\begin{assumption} \label{assumption:f_g}
  Assume that the Jacobian matrix of the local dynamics, $J[\bm f(\bar {\bm x})]$, is symmetric, and the inner-coupling dynamics $\bm g$ is in a linear form with respect to $\bm x$, i.e.,
    \begin{equation*}
      \bm g(\bm x)=a_g\bm x+b_g,
    \end{equation*}
  where $a_g>0$ and $b_g$ are certain scalar constants. \hfs
\end{assumption}

Based on Assumption \ref{assumption:f_g}, we have the following result.
\begin{corollary} \label{corollary:mu_i>0}
  Under Assumption \ref{assumption:f_g}, the pinning controlled network in \eqref{eqn:system_model_pinning} is synchronized at the homogeneous state $\bar {\bm x}$ if and only if
    \begin{equation*}
      \mu_N<-\frac{\lambda_n(J[\bm f(\bar {\bm x})])}{ca_g}.
    \end{equation*}
  \begin{pf}
     Given Assumption \ref{assumption:f_g}, since $\bm g(\bm x)=a_g\bm x+b_g$, it is easy to see that $J[\bm g(\bar {\bm x})]=a_g\bm I$. Since  $J[\bm f(\bar {\bm x})]$ is symmetric, from Lemma \ref{lem:propoerty_eigen_M_N}, the eigenvalues of the matrices $J[\bm f(\bar {\bm x})]+c{\mu_{i}}J[\bm g(\bar {\bm x})]$, $i\in\mathcal V$, are given by:
     \begin{align*}
      &\lambda_{j}(J[\bm f(\bar {\bm x})]+c{\mu_{i}}J[\bm g(\bar {\bm x})])=\lambda_{j}(J[\bm f(\bar {\bm x})])+ca_g{\mu_{i}},\\
      &~~~~~~~~~~~~~~~~~i\in\mathcal V,~j=1,2,...,n.
    \end{align*}
  Therefore, to ensure that all the $N$ matrices $J[\bm f(\bar {\bm x})]+c{\mu_{i}}J[\bm g(\bar {\bm x})]$, $i\in\mathcal V$, are stable, we require all the above $N\times n$ eigenvalues to be negative, which is equivalent to:
  \begin{equation*}
      \lambda_{n}(J[\bm f(\bar {\bm x})])+ca_g{\mu_{N}}<0,~~\text{i.e.,}~~\mu_N<-\frac{\lambda_n(J[\bm f(\bar {\bm x})])}{ca_g}.
    \end{equation*}

    From Theorem~\ref{thm:pinning_controllability}, the above condition will ensure that the pinning controlled network in \eqref{eqn:system_model_pinning} is synchronized at the homogeneous state $\bar {\bm x}$, which completes the proof.
  \end{pf}
\end{corollary}

Intuitively speaking, if we select all the nodes and control them with sufficient large control gains, based on the previous discussions, the eigenvalues of matrix $\mathcal C$ will be positively large enough, while the ones of $\mathcal B$ will be negatively large enough. In such a case, no matter what network coupling structure $\mathcal A$ and the local dynamics $\bm f$ are, the condition in Corollary \ref{corollary:mu_i>0} will always be satisfied. However, the nodes in a complex network are typically huge and it may not be realistic to control them entirely. On the other hand, subject to certain practical constraints, the control gains cannot be arbitrarily large. Therefore, it is worthwhile to consider the optimal pinning node selection problem under reasonable constraints, which will be investigated in the following subsection.

Based on Corollary \ref{corollary:mu_i>0}, the constrained pinning node selection problem becomes properly designing the binary variable $d_i$  with proper control gain $c_i$ in the matrix $\mathcal B$ under certain constraints (e.g., limiting the total number of pinning nodes), such that the largest eigenvalue of $\mathcal B$, $\mu_N$, is less than a threshold $-\lambda_n(J[\bm f(\bar {\bm x})])/ca_g$.

\subsection{Free Selection of Nodes } \label{sec:CaseI}

There are several pinning schemes proposed in the literature and selection priority is given to the nodes with special properties. For example, in \cite{wang2002pinning}, the authors considered a pinning scheme where they first choose the node with the highest degree, and then continue to pin the other nodes in monotonically decreasing order of degrees. The results in \cite{song2012pinning} suggested that we can first divide the the network into components, where each component contains a directed tree, and then select the root nodes of those trees as pinning nodes. However, in practice, the costs to control each node are not identical. Consider, for instance, the advertisement in social networks: In order to achieve a better acknowledgement of the products, the company prefers to choose the ones with more connections (higher degree) in the network, which typically would ask for more endorsement fees. Therefore, the company need to optimize the tradeoffs between the influences and the costs, by selecting the ones with less connections with less costs instead. To take this factor into consideration, define the pinning cost coefficient for node $i$ as $v_i (>0)$, $i\in\mathcal V$ and
\begin{equation*}
\bm v\triangleq[v_1,v_2,...,v_N]'.
 \end{equation*}

On the other hand, the pinning cost for a certain node will also increase with the corresponding control gain $c_i$. As a result, the cost for selecting pinning node $i$ with control gain $c_i$ is given by $d_i v_ic_i$ and the total costs of a certain pinning scheme can be evaluated by the quantity $\sum_{i=1}^Nd_i v_ic_i$.

In such a scenario, the objective of the pinning design problem becomes minimizing the total costs to achieve the network synchronization. The following results reveal that this constrained pinning node selection and control gain design problem can be formulated as a standard convex optimization problem.

\begin{lemma}[\cite{palomar2010convex}]\label{lem:convex_eigenvalue}
If for each $y\in \mathbb D$, the function $w(x, y)$ is convex in $x$, then the function $v$ defined as
\begin{equation*}
  v(x)=\sup_{y\in \mathbb D}w(x,y)
\end{equation*}
 is convex in $x$.
\end{lemma}

\begin{theorem} \label{thm:convex1}
  Let $\mathcal V_{\text{c}}\subseteq\mathcal V$ denote the selectable nodes set for pinning control. The optimal pinning nodes and the corresponding control gains in terms of minimal total costs are given by:
   \begin{equation*}
     d_i=\begin{cases}
       0, & \text{if}~\beta_i= 0,\\
       1, &\text{otherwise},
     \end{cases}
   \end{equation*}
   and
   \begin{equation*}
    c_i=\beta_i c,~~i\in\mathcal V,
   \end{equation*}
   where $\bm \beta\triangleq[\beta_1,\beta_2,...,\beta_N]'$ is the solution to the following convex optimization problem:
\begin{equation} \label{problem1}
  \begin{aligned}
    &\min_{\bm \beta}~ \bm v'\bm \beta\\
    &\text{s.t.} \begin{cases}
       \lambda_{\text{max}}(\mathcal A-{\rm diag}\{\bm \beta \})\leqslant  -\frac{\lambda_n(J[\bm f(\bar {\bm x})])}{ca_g},\\
       \bm \beta\geqslant0,\\
       \beta_i=0,~~i\in\mathcal V\setminus \mathcal V_{\text{c}}.
     \end{cases}
  \end{aligned}
  \end{equation}
  \begin{pf}
Let $\beta_i=c_id_i/c$, it is easy to verify that the largest eigenvalue of the symmetric matrix $\mathcal B$ can be calculated by:
    \begin{align*}
      \mu_N&=\lambda_{\text{max}}(\mathcal B)\\
      &=\sup_{\|y\|_2=1}y'\mathcal By\\
      &=\sup_{\|y\|_2=1}y'(\mathcal {A-C}) y\\
      &=\sup_{\|y\|_2=1}y'(\mathcal {A}-{\rm diag}\{\bm \beta \}) y.
    \end{align*}
    From Lemma \ref{lem:convex_eigenvalue}, $\mu_N=\lambda_{\text{max}}(\mathcal A-{\rm diag}\{\bm \beta \})$ is a convex function of $\bm \beta$. On the other hand, the total costs $\sum_{i=1}^Nd_i v_ic_i$ can be expressed as the quantity $\bm v'\bm \beta$, which is also a convex function of $\bm \beta$. Therefore, the optimization problem in \eqref{problem1} is a convex optimization problem. From Corollary \ref{corollary:mu_i>0}, it is straightforward that the binary variables $d_i$'s  and control gains $c_i$'s obtained from $\beta_i$'s correspond to the optimal pinning node selection and their control gains, respectively, in terms of minimal total costs, which completes the proof.
  \end{pf}
\end{theorem}

Note that the convex optimization problem in \eqref{problem1} can be solved efficiently by various methods and algorithms \cite{palomar2010convex}, e.g., the \texttt{cvx} toolbox (a MATLAB-based modeling system for convex optimization), which also provides built-in function for calculating the largest eigenvalue of a matrix.

\subsection{Free Selection of Nodes: Identical Control Gains} \label{subsec:free_identical}

Now we consider a special case when identical control gains are adopted. Without loss of generality, we assume that $c_i=\bar c$, $\forall i\in\mathcal V$. The objective in such a scenario becomes minimizing the total cost through purely node selection to achieve the synchronization. In particular, when the pinning cost coefficients $v_i$'s for all nodes are the same, the problem becomes minimizing the total number of pinning nodes to achieve network synchronization.

As a result, we obtain Theorem \ref{thm:BLP1} directly from Theorem~\ref{thm:convex1} by adding binary constraints on $\beta$.

\begin{theorem} \label{thm:BLP1}
  Given an identical control gain $c_i=\bar c$, $\forall i\in\mathcal V$, the optimal pinning node selection solution in terms of minimal total costs is given by
  \begin{equation*}
     \bm d=\bm \beta,
   \end{equation*}
   where $\bm \beta=[\beta_1,\beta_2,...,\beta_N]'$ is the binary solution to the following problem:
   \begin{equation} \label{problem2}
  \begin{aligned}
    &\min_{\bm \beta}~ \bm v'\bm \beta\\
    &\text{s.t.} \begin{cases}
       \lambda_{\text{max}}(\mathcal A-\bar c/c\cdot{\rm diag}\{\bm \beta \})\leqslant  -\frac{\lambda_n(J[\bm f(\bar {\bm x})])}{ca_g},\\
       \bm \beta\in\{0,1\}^N,\\
       \beta_i=0,~~i\in\mathcal V\setminus \mathcal V_{\text{c}}.
     \end{cases}
  \end{aligned}
  \end{equation}
\end{theorem}

Though Theorem \ref{thm:BLP1} can be regarded as a special case of Theorem \ref{thm:convex1}, the problem in \eqref{problem2} is more complicated than the one in \eqref{problem1}. Due to the binary constraint on $\bm \beta$ ($\bm \beta\in\{0,1\}^N$, which is not a convex set), the problem in \eqref{problem2} (the so-called ``Binary Integer Programming (BIP)" problem) is no longer a convex optimization problem, and thus cannot be solved by standard convex optimization techniques. In fact, standard programming models as in  \eqref{problem1} have continuous decision variables and fractional solutions, which are sometimes not realistic for problem in \eqref{problem2} \cite{garfinkel1972integer}. Noticing that  $\bm \beta\in\{0,1\}^N$ has $2^N$ possible values, a direct enumeration approach to obtain the optimal nodes selection solution is therefore computationally intractable when the network scale $N$ is large.

Among the various algorithms in the literature \cite{geoffrion1972integer}, the Branch-and-Bound (B\&B) method is the most popular one for solving large scale NP-hard combinatorial optimization problems \cite{clausen1999branch}. Though the entire solution space may be searched in the worst-case scenario, the utilization of bounds generated from the current best solution generally helps reduce the computation process and  a smaller solution space is searched instead.

To be more specific, denote $\mathcal V_{\text{sub}},\mathcal V_{\text{fix}}\subseteq\mathcal V$ as two certain sub-index sets of $\mathcal V$, and $\bm \beta(\mathcal V_{\text{sub}}),\bm \beta(\mathcal V_{\text{fix}})$ as the corresponding collections of elements from $\bm \beta$, respectively, and we can summarize this iterative approach in Algorithm \ref{algm:BB}.

\begin{algorithm} [ht]
\caption{B\&B method for solving the BIP in \eqref{problem2}} \label{algm:BB}
\begin{algorithmic}[1]
\State Initialization: set $ incu=+\infty$, $\mathcal V_{\text{sub}}=\mathcal V$, $\mathcal V_{\text{fix}}=\emptyset$, and $\bm \beta(\mathcal V_{\text{fix}})=\emptyset$;
\Function{$B\&B$}{$\mathcal V_{\text{sub}},\mathcal V_{\text{fix}}, \bm \beta(\mathcal V_{\text{fix}}), incu$}
\State Solve the following convex problem:
\begin{equation*}
  \begin{aligned}
    &\min_{\bm \beta}~ \bm v'\bm \beta\\
    &\text{s.t.} \begin{cases}
       \lambda_{\text{max}}(\mathcal A-\bar c/c\cdot{\rm diag}\{\bm \beta \})\leqslant  -\frac{\lambda_n(J[\bm f(\bar {\bm x})])}{ca_g},\\
       \bm \beta(\mathcal V_{\text{sub}})\in[0,1]^{|V_{\text{sub}}|},\\
       \bm \beta(\mathcal V\setminus\mathcal V_{\text{sub}})=\bm \beta(\mathcal V_{\text{fix}}),\\
       \beta_i=0,~~i\in\mathcal V\setminus \mathcal V_{\text{c}};
     \end{cases}
  \end{aligned}
  \end{equation*}
\If {the solution $\bm \beta_{\text{sol}}$ exists and $\bm v'\bm \beta_{\text{sol}}<incu$}
\If{the solution $\bm \beta_{\text{sol}}$ is in binary form}
\State \Return $\bm \beta_{\text{sol}}$;
\Else
\State Choose $k\in\mathcal V_{\text{sub}}$ and define:
 \begin{equation*}
 \begin{cases}
    \tilde {\mathcal V}_{\text{sub}}=\mathcal V_{\text{sub}}\setminus\{k\},\\
    \tilde{\mathcal V}_{\text{fix}}=\mathcal V_{\text{fix}}\cup\{k\},\\
    \tilde{\bm \beta}_1(\mathcal V_{\text{fix}})=\bm \beta(\mathcal V_{\text{fix}})\cup\{\beta_k=1\},\\
    \tilde{\bm \beta}_0(\mathcal V_{\text{fix}})=\bm \beta(\mathcal V_{\text{fix}})\cup\{\beta_k=0\};
     \end{cases}
 \end{equation*}
\State  $\bm \beta_{\text{sol1}}=$B\&B$(\tilde {\mathcal V}_{\text{sub}},\tilde {\mathcal V}_{\text{fix}},\tilde {\mathcal V}_{\text{sub}},\tilde {\bm \beta}_1(\mathcal V_{\text{fix}}), incu)$;
\State  $\bm \beta_{\text{sol0}}=$B\&B$(\tilde {\mathcal V}_{\text{sub}},\tilde {\mathcal V}_{\text{fix}},\tilde {\mathcal V}_{\text{sub}},\tilde {\bm \beta}_0(\mathcal V_{\text{fix}}), incu)$;
\State  $\tilde{\bm \beta}_{\text{sol}}=\arg_{\{\bm \beta_{\text{sol1}},\bm \beta_{\text{sol0}}\}}\min\{\bm v'\bm \beta_{\text{sol1}},\bm v'\bm \beta_{\text{sol0}}\}$;
\If{$\bm v'\tilde{\bm \beta}_{\text{sol}}< incu$}
\State  $incu=\bm v'\tilde{\bm \beta}_{\text{sol}}$;
\State \Return $\tilde{\bm \beta}_{\text{sol}}$;
\Else
\State \Return $null$;
\EndIf

\EndIf
\Else
\State \Return $null$;
\EndIf
\EndFunction
\end{algorithmic}
\end{algorithm}

\begin{remark}
  The computational complexity of the branch and bound method depends on the specific problem structure and the branch nodes selection protocol, which is studied comprehensively in the literature, e.g., \cite{zhang1996branch,thakoor2009computation}. We will not investigate this in details in this paper.\hfs
\end{remark}

\subsection{Constrained Number of Pinning Nodes}

The situation in Section \ref{sec:CaseI} assumes a free selection of nodes. However, in several applications, there may exist constraint on the total number of pinning nodes. We will investigate the pinning node selection and control gain design problem under such constraint in this part.

Denote the maximal total number of pinning nodes as $N_{\text{total}}$. Based on Theorem \ref{thm:convex1}, by adding a cardinality constraint on $\bm \beta$, we can readily obtain the following result.
\begin{theorem} \label{thm:convex_cardinality}
  Given the maximal total number of pinning nodes $N_{\text{total}}$, the optimal pinning nodes and the corresponding control gains in terms of minimal total costs are given by:
   \begin{equation*}
     d_i=\begin{cases}
       0, & \text{if}~\beta_i= 0,\\
       1, &\text{otherwise},
     \end{cases}
   \end{equation*}
   and
   \begin{equation*}
    c_i=\beta_ic,~~i\in\mathcal V,
   \end{equation*}
   where $\bm \beta\triangleq[\beta_1,\beta_2,...,\beta_N]'$ is the solution to the following convex optimization problem:
\begin{equation} \label{problem3}
  \begin{aligned}
    &\min_{\bm \beta}~ \bm v'\bm \beta\\
    &\text{s.t.} \begin{cases}
       \lambda_{\text{max}}(\mathcal A-{\rm diag}\{\bm \beta \})\leqslant  -\frac{\lambda_n(J[\bm f(\bar {\bm x})])}{ca_g},\\
       \bm \beta\geqslant0,\\
       \textbf{card}(\bm\beta)\leqslant N_{\text{total}}.
     \end{cases}
  \end{aligned}
  \end{equation} \hfs
\end{theorem}

It is easy to verify that $\textbf{card}(x)$ is quasiconcave on $\mathbb R^n_+$, which make the problem in \eqref{problem3} a convex-cardinality problem \cite{boyd2004convex}. Similar to the problem in \eqref{problem2}, standard convex optimization techniques are not suitable. However, notice that if we fix the sparsity pattern of $\bm\beta$, i.e., which elements are zero or non-zero, the problem in \eqref{problem3} becomes a convex problem. We can divide the convex-cardinality problem into convex sub-problems according to all the possible sparsity patterns, which is  a NP-hard problem. Fortunately, it can be solved by the branch and bound method that we discussed in the previous part. The procedure is similar to the one in Algorithm \ref{algm:BB} and thus we omit it here.

\section{Stackelberg Security Game Analysis}

In this section, we continue the analysis of the pinning node selection problem from a security perspective.

\subsection{Defender and Attacker Models} \label{subsec:d_and_a_model}

Without loss of generality, we develop the following analysis on the basis of free selection of nodes with identical control gains as in Section \ref{subsec:free_identical}. The extensions to other cases are similar. Assume that a  pre-determined selection of nodes (represented by the set $\mathcal V_{\text{pin}}\subseteq\mathcal V$) to apply feedback controllers with an identical control gain, i.e.,
\begin{equation*}
\bm d=\bm \beta ({\mathcal V_\text{pin}})\triangleq\bm \beta_{\text{pin}}=\big[{\beta}_{\text{pin}}^{(1)},{\beta}_{\text{pin}}^{(2)},...,{\beta}_{\text{pin}}^{(N)}\big]'\in\{0,1\}^N.
\end{equation*}
 Notice that for a faster convergence rate or the redundancy concern \cite{wang2002pinning}, the pinning control scheme $\bm \beta_{\text{pin}}$ is assumed to include extra nodes than the optimal solution to the problem in ~\eqref{problem2} and can guarantee the synchronization of the pinning controlled network at a homogeneous state in the case without an attacker. As stated before, to ensure the normal operation of the complex networks and defend malicious attacks, one straightforward strategy for the defender is to secure the nodes in the networks. Under the limited defensive budget, we assume that the defender's strategy is given by a vector:
\begin{equation*}
  \bm {\pi}_d\triangleq\big[\pi_d^{(1)},\pi_d^{(2)},...,\pi_d^{(N)}\big]',
\end{equation*}
where $\pi_d^{(i)}$ is the defense budget allocated to the node $i$. The total cost for establishing such a defense strategy can be written as:
\begin{equation*}
  \mathcal U_d\triangleq \bm 1\cdot\bm {\pi}_d.
\end{equation*}

Further assume that the associated cost for compromising the node $i$, denoted by $\sigma^{(i)}$, is linear with its allocated defense investment $\pi_d^{(i)}$, i.e., $\sigma^{(i)}=\kappa^{(i)}\pi_d^{(i)}$, where the coefficient $\kappa^{(i)}>0$ represents how difficult for the attacker to compromise the node $i$. For the ease of notation, we denote:
\begin{equation*}
  \bm \sigma \triangleq\big[\sigma^{(1)},\sigma^{(2)},...,\sigma^{(N)}\big]',
\end{equation*}
and
 \begin{equation*}
  \bm {\kappa}\triangleq\big[\kappa^{(1)},\kappa^{(2)},...,\kappa^{(N)}\big]'.
\end{equation*}

On the other hand, suppose that the attacker aims to compromise the nodes in the complex networks with pinning attacking index vector denoted as
\begin{equation*}
  \bm {\beta}_{\text{attack}}\triangleq\big[{\beta}_{\text{attack}}^{(1)},{\beta}_{\text{attack}}^{(2)},...,{\beta}_{\text{attack}}^{(N)}\big]'\in\{0,1\}^N,
\end{equation*}
where ${\beta}_{\text{attack}}^{(i)}=1$ or $0$ indicates whether the attacker will compromise the node $i$ or not, respectively. Then, given the defender's resource allocation, the cost for compromising the target nodes indexed by $\bm {\beta}_{\text{attack}}$  can be written as:
\begin{equation*}
  \mathcal U_a\triangleq\bm \sigma\cdot\bm {\beta}_{\text{attack}} =(\bm {\kappa}\circ\bm {\pi}_d)\cdot\bm {\beta}_{\text{attack}},
\end{equation*}
where the operator $\circ$ denotes the Hadamard product (entrywise product).

In the following discussion, the superscript ``$\tilde\cdot$'' is used to denote the associated quantities under attacks. Given the pinning control scheme $\bm\beta_{\text{pin}}$ and the pinning attack scheme $\bm  {\beta}_{\text{attack}}$, we have $\tilde d_i={\beta}_{\text{pin}}^{(i)}(1-{\beta}_{\text{attack}}^{(i)})$ and thus $\bm{\tilde d}=\bm {\beta}_{\text{pin}}\circ(\bm 1-\bm {\beta}_{\text{attack}})$. Therefore, the pinning controlled network in \eqref{eqn:system_model_pinning} under attacks can be written as:
\begin{equation} \label{eqn:netwrok_under_attack}
  \dot{\bm x}_i=\bm f(\bm x_i)+c\sum_{j=1}^Na_{ij}\bm g(\bm x_j)+{\beta}_{\text{pin}}^{(i)}(1-{\beta}_{\text{attack}}^{(i)})\bm u_i,~~i\in\mathcal V.
\end{equation}

Now the matrices $\mathcal C$ and $\mathcal B$ in \eqref{def:mathcal_C} and \eqref{def:mathcal_B} can be written as:
\begin{equation*}
\tilde{\mathcal C}\triangleq \bar c/c\cdot{\rm diag}\{\tilde{\bm d}\}\in\mathbb R^{N\times N},
\end{equation*}
and
\begin{equation*}
 \tilde{\mathcal B} \triangleq \mathcal {A-\tilde C}\in\mathbb R^{N\times N}.
\end{equation*}

From Corollary~\ref{corollary:mu_i>0}, under  pinning attacks, the pinning controlled network in \eqref{eqn:netwrok_under_attack} is synchronized at the state $\bar {\bm x}$ if and only if:
\begin{equation*}
\lambda_{\text{max}}(\tilde{\mathcal B})\leqslant  -\frac{\lambda_n(J[\bm f(\bar {\bm x})])}{ca_g}.
\end{equation*}
The attacker compromises a certain portion of nodes to affect the synchronization of the
pinning controlled network, so that
\begin{equation*}
\lambda_{\text{max}}(\tilde{\mathcal B})\geqslant  -\frac{\lambda_n(J[\bm f(\bar {\bm x})])}{ca_g},
\end{equation*}
i.e.,
\begin{equation}\label{eqn:objective_attacker}
\lambda_{\text{max}}({\mathcal A}-\bar c/c\cdot{\rm diag}\{{\bm {\beta}_{\text{pin}}\circ(\bm 1-\bm {\beta}_{\text{attack}})}\})\geqslant -\frac{\lambda_n(J[\bm f(\bar {\bm x})])}{ca_g}.
\end{equation}

On the other hand, to protect the network, the defender aims to allocate the secure resources among the pinning nodes efficiently to increase the attacking cost of the attacker. Based on the proposed defender and attacker models, we can summarize  the elements of such a two-player Stackelberg security game as follows:
\begin{itemize}
  \item \textit{Players}: the defender and the attacker;
  \item \textit{Actions}: the action of the defender is represented by its secure resources allocation strategy $\bm {\pi}_d$; similarly, the action of the attacker corresponds to its the pinning attack scheme $\bm  {\beta}_{\text{attack}}$;
  \item \textit{Payoffs}: given the condition in \eqref{eqn:objective_attacker} satisfied, the attacker wants to minimize the cost for launching such an attack, namely, $\mathcal U_a$, and we define the payoff for the attacker as:
      \begin{equation} \label{eqn:payoff_R_a}
        \mathcal R_a(\bm  {\beta}_{\text{attack}},\bm \pi_d)\triangleq-\mathcal U_a;
       \end{equation}
      the defender aims to increase the difficulty of launching attacks for the attacker, in terms of $\mathcal U_a$, while reducing its own  defensive in terms of $\mathcal U_d$. Therefore, we propose the payoff for the defender as:
      \begin{equation} \label{eqn:payoff_R_d}
        \mathcal R_d(\bm  {\beta}_{\text{attack}},\bm \pi_d)\triangleq \mathcal U_a-\eta \mathcal U_d,
       \end{equation}
       where $\eta$ is a weighting parameter. Clearly, both players want to maximize their own payoffs.\hfs
\end{itemize}

Traditionally, the security game analysis is in terms of the so-called ``Nash equilibrium" \cite{nash1951game}, where the actions from players are chosen {\it simultaneously} without knowing the action from their opponent in advance, see \cite{agah2004game,li2015jamming}. The situation in this paper is different, where both sides make their decisions {\it sequentially}: it is more reasonable that the defender decides the defense budget allocation first, based on its prediction of the possible reaction from the attacker, then the attacker observes the existing protection situation and select the corresponding nodes to compromise. In the following parts, we use a leader-follower framework (the so-called ``Stackelberg game" \cite{korzhyk2011stackelberg}) to model this security game.

\subsection{Stackelberg Security Game}

Based on the proposed defender/attacker models and the framework of the two-player Stackelberg security game, we have the following results.

\begin{theorem} \label{thm:Solution_stackelberg}
The solutions to the  Stackelberg security game described in Section~\ref{subsec:d_and_a_model} are given by solving:
\begin{equation} \label{eqn:Stackel_best_response_condition_equal}
\max_{\bm \pi_d}\min_{\bm  {\beta}_{\text{attack}}} ~\mathcal R_d(\bm  {\beta}_{\text{attack}},\bm \pi_d),
\end{equation}
given the condition in \eqref{eqn:objective_attacker} satisfied. In particular, the optimal resource allocation for the defender $\bm {\pi}_d^\star$ can be obtained by solving the following linear programming (LP) problem:
\begin{equation} \label{eqn:problem_pi_d_star_LP}
\begin{aligned}
    &\min_{\bm {\pi}_d,\epsilon\in\mathbb R}  ~\epsilon\\
    & \text{s.t.} \begin{cases}
       \mathcal M \bm {\pi}_d\geqslant-\epsilon \bm 1,\\
       \bm {\pi}_d\geqslant \bm 0,
     \end{cases}
\end{aligned}
\end{equation}
where the matrix $\mathcal M$ is given in \eqref{def:mathcal_M}. The optimal pinning attacking node selection for the attacker $\bm  {\beta}_{\text{attack}}^\star$ is obtained by solving:
\begin{equation*}
 \bm  {\beta}_{\text{attack}}^\star=\arg\min_{ \bm  {\beta}_{\text{attack}} \in \{\bm {\beta}_{\text{attack},i}\}_{i=1}^{t_0}}(\bm {\kappa}\circ\bm {\pi}_d^\star)\cdot\bm {\beta}_{\text{attack}}.
\end{equation*}

\begin{pf}
We first call for the concept of best response, which refers to the action that produces the maximal payoff for a player, while taking other players' actions as given \cite{gibbons1992primer}. Specifically, the best responses for the defender and the attacker are defined as:
\begin{equation} \label{eqn:Best_response_B_d}
  \mathcal Y_d(\bm  {\beta}_{\text{attack}})\triangleq\arg\max_{\bm \pi_d} \mathcal R_d(\bm  {\beta}_{\text{attack}},\bm \pi_d),
\end{equation}
and
\begin{equation} \label{eqn:Best_response_B_a}
  \mathcal Y_a(\bm \pi_d)\triangleq\arg\max_{\bm  {\beta}_{\text{attack}}} \mathcal R_a(\bm  {\beta}_{\text{attack}},\bm \pi_d),
\end{equation}
respectively.

Since the decisions in the current Stackelberg security game are sequentially made, given a certain $\bm \pi_d$, the attacker will choose  $\bm  {\beta}_{\text{attack}}=\mathcal Y_a(\bm {\pi}_d)$. Therefore, the defender will choose $\bm \pi_d$ to maximize its payoff given $\bm  {\beta}_{\text{attack}}=\mathcal Y_a(\bm {\pi}_d)$ by solving:
\begin{equation} \label{eqn:Stackel_best_response_condition_d}
  \bm \pi_d^\star=\mathcal Y_d\big(\mathcal Y_a(\bm {\pi}_d^\star)\big),
\end{equation}
The attacker observes $\bm \pi_d^\star$ and then obtain its optimal strategy by solving:
\begin{equation} \label{eqn:Stackel_best_response_condition_a}
 \bm  {\beta}_{\text{attack}}^\star=\mathcal Y_a(\bm {\pi}_d^\star).
\end{equation}

Based on \eqref{eqn:payoff_R_a} and \eqref{eqn:payoff_R_d}, since the term $-\eta\mathcal U_d$ is independent of $\bm  {\beta}_{\text{attack}}$,  \eqref{eqn:Best_response_B_a} can be re-written as:
\begin{equation*}
  \mathcal Y_a(\bm {\pi}_d)=\arg\max_{\bm  {\beta}_{\text{attack}}} \mathcal R_a(\bm \gamma_a,\bm \pi_d)=\arg\min_{\bm  {\beta}_{\text{attack}}} \mathcal R_d(\bm \gamma_a,\bm \pi_d).
\end{equation*}

Therefore, from \eqref{eqn:Stackel_best_response_condition_d} and \eqref{eqn:Stackel_best_response_condition_a}, the optimal solutions for both sides in this sequential decision-making Stackelberg security game can be obtained by solving:
\begin{equation*}
  \max_{\bm \pi_d}\min_{\bm  {\beta}_{\text{attack}}} \mathcal R_d(\bm  {\beta}_{\text{attack}},\bm \pi_d),
\end{equation*}
given the condition in \eqref{eqn:objective_attacker} satisfied.

Clearly, the attacker only needs to compromise the nodes belonging to the pre-determined set $\mathcal V_{\text{pin}}$, to which the feedback controllers are applied. Then we have:
\begin{equation} \label{eqn:beta_attack_0_vpin}
  {\beta}_{\text{attack}}^{(i)}=0,~\text{if}~i\in\mathcal V\setminus \mathcal V_{\text{pin}}.
\end{equation}
As a result, the condition in \eqref{eqn:objective_attacker} can be re-written as:
\begin{equation} \label{eqn:objective_attacker_new}
\begin{cases}
       \lambda_{\text{max}}(\mathcal A-\bar c/c\cdot{\rm diag}\{\bm {\beta}_{\text{pin}} \}+\bar c/c\cdot{\rm diag}\{\bm {\beta}_{\text{attack}} \})\\
       ~~~~~~~~~~~~~~~~~~~~~~~~~~~~~~~~~~~~~~~~~~~~~\geqslant -\frac{\lambda_n(J[\bm f(\bar {\bm x})])}{ca_g},\\
         {\beta}_{\text{attack}}^{(i)}=0,~\text{if}~i\in\mathcal V\setminus \mathcal V_{\text{pin}}.
     \end{cases}
\end{equation}

Clearly, given the set $\mathcal V_{\text{pin}}$, the total number of the possible values $\bm {\beta}_{\text{attack}}$ can take is given by $2^{|\mathcal V_{\text{pin}}| }$. Suppose that among these $2^{|\mathcal V_{\text{pin}}| }$ possible pinning attacking node selections, there are $t_0$ values for $\bm {\beta}_{\text{attack}}$ to satisfy the first condition in \eqref{eqn:objective_attacker_new}. We denote them by $\{\bm {\beta}_{\text{attack},i}\}_{i=1}^{t_0}$.

Similarly, the defender will only allocate protective resources to the nodes in  $\mathcal V_{\text{pin}}$, i.e.,
\begin{equation*}
  {\pi}_{d}^{(i)}=0,~\text{if}~i\in\mathcal V\setminus \mathcal V_{\text{pin}}.
\end{equation*}

By summarizing the previous discussions and from Wald's maximin model \cite{wald1950statistical}, solving \eqref{eqn:Stackel_best_response_condition_equal} becomes an equivalent mathematical programming (MP) problem:
\begin{align*}
  &\max_{\bm \pi_d}\min_{\bm  {\beta}_{\text{attack}}} \mathcal R_d(\bm  {\beta}_{\text{attack}},\bm \pi_d)\\
= &\max_{\bm \pi_d}\min_{1\leqslant i \leqslant t_0 }  \big\{\mathcal R_d(\bm  {\beta}_{\text{attack},i},\bm \pi_d), \forall i=1,2,...,  t_0  \big\}\\ \notag
=&\max_{\bm \pi_d,\epsilon\in\mathbb R} \big\{  \epsilon  \big|  \epsilon \leqslant\mathcal R_d(\bm  {\beta}_{\text{attack},i},\bm \pi_d),~ \forall i=1,2,...,t_0 \big\}\\ \notag
=&\max_{\bm \pi_d,\epsilon\in\mathbb R} \big\{  \epsilon  \big|  \epsilon \leqslant \big[(\bm {\kappa}\circ\bm {\beta}_{\text{attack},i})-\eta\bm 1\big]\cdot\bm {\pi}_d,\\  \notag
&~~~~~~~~~~~~~~~~~~~~~~~~~~~~~~~~~~~~~~\forall i=1,2,..., t_0 \big\}\\ \notag
=&\max_{\bm \pi_d,\epsilon\in\mathbb R} \big\{  \epsilon  \big|  \epsilon \bm 1 \leqslant \mathcal M\bm {\pi}_d\big\},
\end{align*}
where
\begin{equation} \label{def:mathcal_M}
  \mathcal{M} \triangleq \big[ \mathcal{M}[1]',\mathcal{M}[2]',...,\mathcal{M}[ t_0 ]'\big]'\in\mathbb R^{ t_0 \times N},
\end{equation}
with
\begin{equation*}
    \mathcal{M}[i]= \big[(\bm {\kappa}\circ\bm {\beta}_{\text{attack},i})-\eta\bm 1\big]',
\end{equation*}
as the $i$-th row of $\mathcal{M}$.

Combining with condition in \eqref{eqn:objective_attacker_new}, we can formally re-write the above optimization problem in the following LP problem:
  \begin{align*}
    &\min_{\bm {\pi}_d,\epsilon\in\mathbb R}  ~\epsilon\\
    & \text{s.t.} \begin{cases}
       \mathcal M \bm {\pi}_d\geqslant-\epsilon \bm 1,\\
       \bm {\pi}_d\geqslant \bm 0,
     \end{cases}
  \end{align*}

After we obtain the solution $\bm {\pi}_d^\star$ for the defender, from \eqref{eqn:Stackel_best_response_condition_a}, the optimal node selection for compromising is readily obtained by:
\begin{equation*}
 \bm  {\beta}_{\text{attack}}^\star=\arg\min_{ \bm  {\beta}_{\text{attack}} \in \{\bm {\beta}_{\text{attack},i}\}_{i=1}^{t_0}}(\bm {\kappa}\circ\bm {\pi}_d^\star)\cdot\bm {\beta}_{\text{attack}},
\end{equation*}
which completes the proof.
\end{pf}
\end{theorem}

\subsection{Discussions on Fixed Budget Constraints}

In the previous subsection, Theorem~\ref{thm:Solution_stackelberg} provides general solutions to the Stackelberg security problem where the resource constraints for both sides are imposed by a penalty term in their objective functions. However, in certain cases, the attacker or the defender may be constrained to fixed budgets. We will study these cases briefly in this section.

First we consider the scenario when the defender has a fixed total resource limit, denoted as $\overline\Omega_d$. In such a case, the defender will not put a penalty term on the total defense cost, namely, $\eta=0$; instead, this term is replaced by a constraint in the optimization problem. Now the problem in~\eqref{eqn:problem_pi_d_star_LP} can be re-cast as follows.
\begin{proposition} \label{prop:Pi_d}
  The solution of $\bm {\pi}_d^\star$ for a fixed defender budget $\overline\Omega_d$ is given by solving the following LP problem with $\eta=0$:
\begin{equation} \label{eqn:problem_pi_d_star_LP_Pi_D}
\begin{aligned}
    &\min_{\bm {\pi}_d,\epsilon\in\mathbb R}  ~\epsilon\\
    & \text{s.t.} \begin{cases}
       \mathcal M \bm {\pi}_d\geqslant-\epsilon \bm 1,\\
       \bm {\pi}_d\geqslant \bm 0,\\
       \bm 1\cdot \bm {\pi}_d\leqslant \overline\Omega_d
     \end{cases}
\end{aligned}
\end{equation}\hfs
\end{proposition}

When a fixed budget (denoted as $\overline\Omega_a$) goes to the attacker for compromising nodes, the objective of the defender becomes minimizing the total defense budget while keeping the minimal cost of launching a successful pinning attack beyond the attacker's resource constraint $\overline\Omega_a$, which is summarized in as follows.

\begin{proposition}\label{prop:Pi_a}
  The solution of $\bm {\pi}_d^\star$ for a fixed attacking budget $\overline\Omega_a$ is given by solving the following LP problem with $\eta=0$:
\begin{equation} \label{eqn:problem_pi_d_star_LP_Pi_A}
  \begin{aligned}
    &\min_{\bm {\pi}_d}  ~\bm 1\cdot \bm {\pi}_d\\
    & \text{s.t.} \begin{cases}
\mathcal{ M} \bm {\pi}_d\geqslant   \overline\Omega_a\bm 1,\\
    \bm {\pi}_d\geqslant \bm 0,
     \end{cases}
  \end{aligned}
\end{equation}
\end{proposition}

From the aforementioned discussions, by allocating the defense resource according to the solution to the problem in \eqref{eqn:problem_pi_d_star_LP_Pi_A}, the defender can protect the network from the pinning attack launched by the attacker with fixed budget $\overline\Omega_a$. On the other hand, when the solution $\bm {\pi}_d^\star$ is beyond the defender's fixed budget, i.e., $\bm 1\cdot \bm {\pi}_d^\star> \overline\Omega_d$, the defender is not capable to prevent the network from the pinning attacks.

\section{Numerical Examples}

In this section, we will illustrate our main results using several simple numerical examples.

\begin{figure}[ht]
  \centering
  \includegraphics[width=7cm]{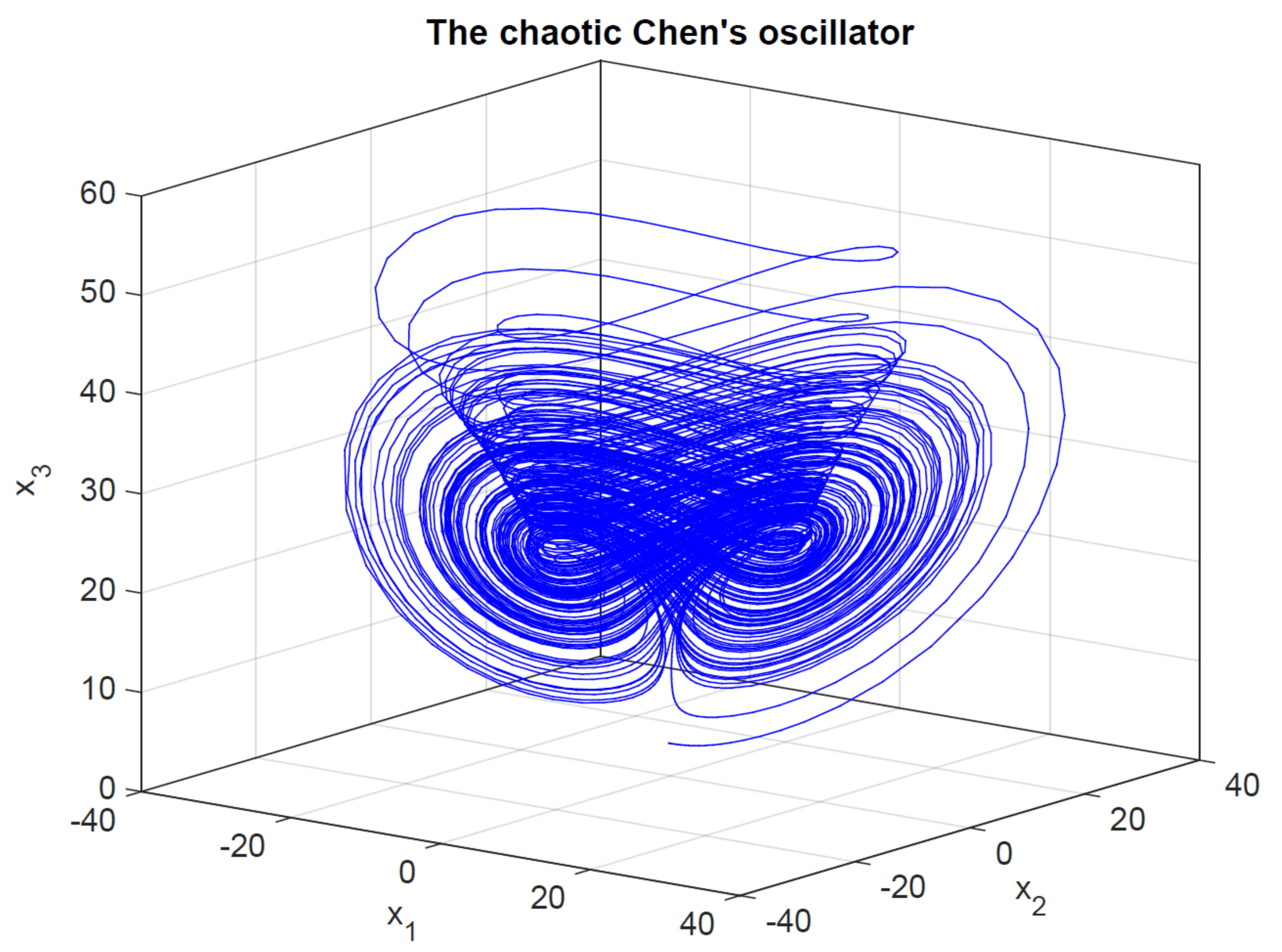}
  \caption{An illustration of the chaotic Chen's oscillator with $x_1(0)=-1$, $x_2(0)=1$, $x_3(0)=3$.} \label{fig:chen}
\end{figure}

The chaotic Chen's oscillator  \cite{chen1999yet} is typically used as the dynamic model for a single node \cite{xiang2007pinning}, which is given by:
\begin{equation*}
\left(\begin{array}{c}
  \dot x_1 \\
  \dot x_2 \\
  \dot x_3
\end{array}\right)=\left(\begin{array}{c}
 \alpha(x_2- x_1) \\
  (\gamma-\alpha)x_1-x_1x_3+\gamma x_2 \\
  x_1x_2-\beta x_3
\end{array}\right),
\end{equation*}
with $\alpha=35$, $\beta=3$, $\gamma=28$. An illustration of the chaotic Chen's oscillator is shown in Figure~\ref{fig:chen}.

\begin{figure}[ht]

  \centering

\usetikzlibrary{shapes}
\begin{tikzpicture}[scale=1.2]

\draw (0,0) circle (0.15);
\node at (0,0){4};

\draw (-1.5,0) circle (0.15);
\node at (-1.5,0){2};
\draw (-0.15,0) -- (-1.35,0);

\draw (1.5,0) circle (0.15);
\node at (1.5,0){6};
\draw (0.15,0) -- (1.35,0);

\draw (60:1.5) circle (0.15);
\node at (60:1.5) {5};
\draw (60:0.15) -- (60:1.35);

\draw (-60:1.5) circle (0.15);
\node at (-60:1.5) {7};
\draw (-60:0.15) -- (-60:1.35);

\draw (-120:1.5) circle (0.15);
\node at (-120:1.5) {3};
\draw (-120:0.15) -- (-120:1.35);

\draw (120:1.5) circle (0.15);
\node at (120:1.5) {1};
\draw (120:0.15) -- (120:1.35);

\draw (0.825,1.16913) -- (1.425, 0.1299);

\draw (0.825,-1.16913) -- (1.425, -0.1299);

\draw (2.25,1.299) circle (0.15);
\node at  (2.25,1.299) {8};
\draw (0.9,1.299) -- (2.1,1.299);

\draw (2.25,-1.299) circle (0.15);
\node at  (2.25,-1.299) {9};
\draw (0.9,-1.299) -- (2.1,-1.299);

\end{tikzpicture}

  \caption{The network topology for the simulation.} \label{fig:topology}

\end{figure}
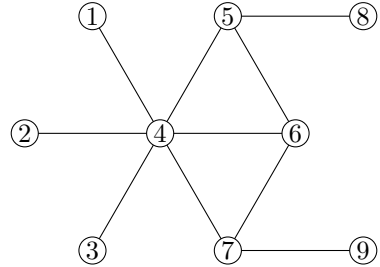

In our simulation, assume that $a_g=1$, $b_g=0$, $c=10$, $N=9$. The  state equations in~\eqref{eqn:system_model} for the complex network with nodes modelled by the Chen's oscillator  can be written as:
\begin{equation*}
\left(\begin{array}{c}
  \dot x_{i,1} \\
  \dot x_{i,2} \\
  \dot x_{i,3}
\end{array}\right)=\left(\begin{array}{c}
 \alpha(x_{i,2}- x_{i,1}) +c\sum_{j=1}^Na_{ij} x_{j,1}\\
  (\gamma-\alpha)x_{i,1}-x_{i,1}x_{i,3}+\gamma x_{i,2} \\
  ~~~~~~~~~~~~~~~~~~~~~+c\sum_{j=1}^Na_{ij} x_{j,2}\\
  x_{i,1}x_{i,2}-\beta x_{i,3}+c\sum_{j=1}^Na_{ij} x_{j,3}
\end{array}\right),
\end{equation*}
where $i=1,2,...,9$, with the topology depicted as in Figure~\ref{fig:topology}.

\begin{figure}[ht]

\centering

\subfigure[The node states evolution in the complex network without pinning control.]
{
\begin{minipage}[c]{0.5\textwidth}
\centering
\includegraphics[width=8.5cm]{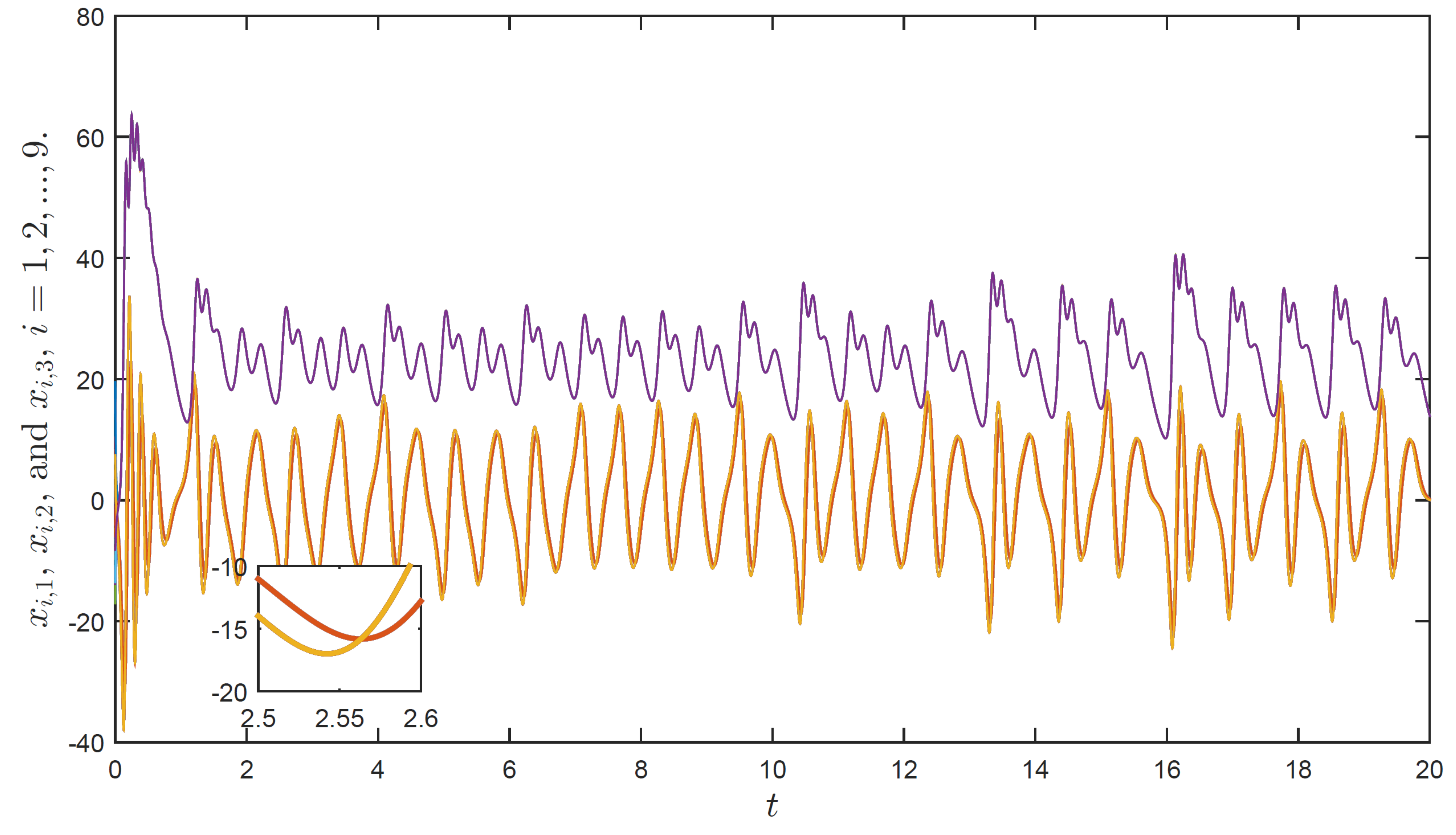}
\end{minipage}%
}

\centering

\subfigure[The node state evolution in the complex network under pinning control scheme $\bm \beta_1$.]
{
\begin{minipage}[c]{0.5\textwidth}
\centering
\includegraphics[width=8.5cm]{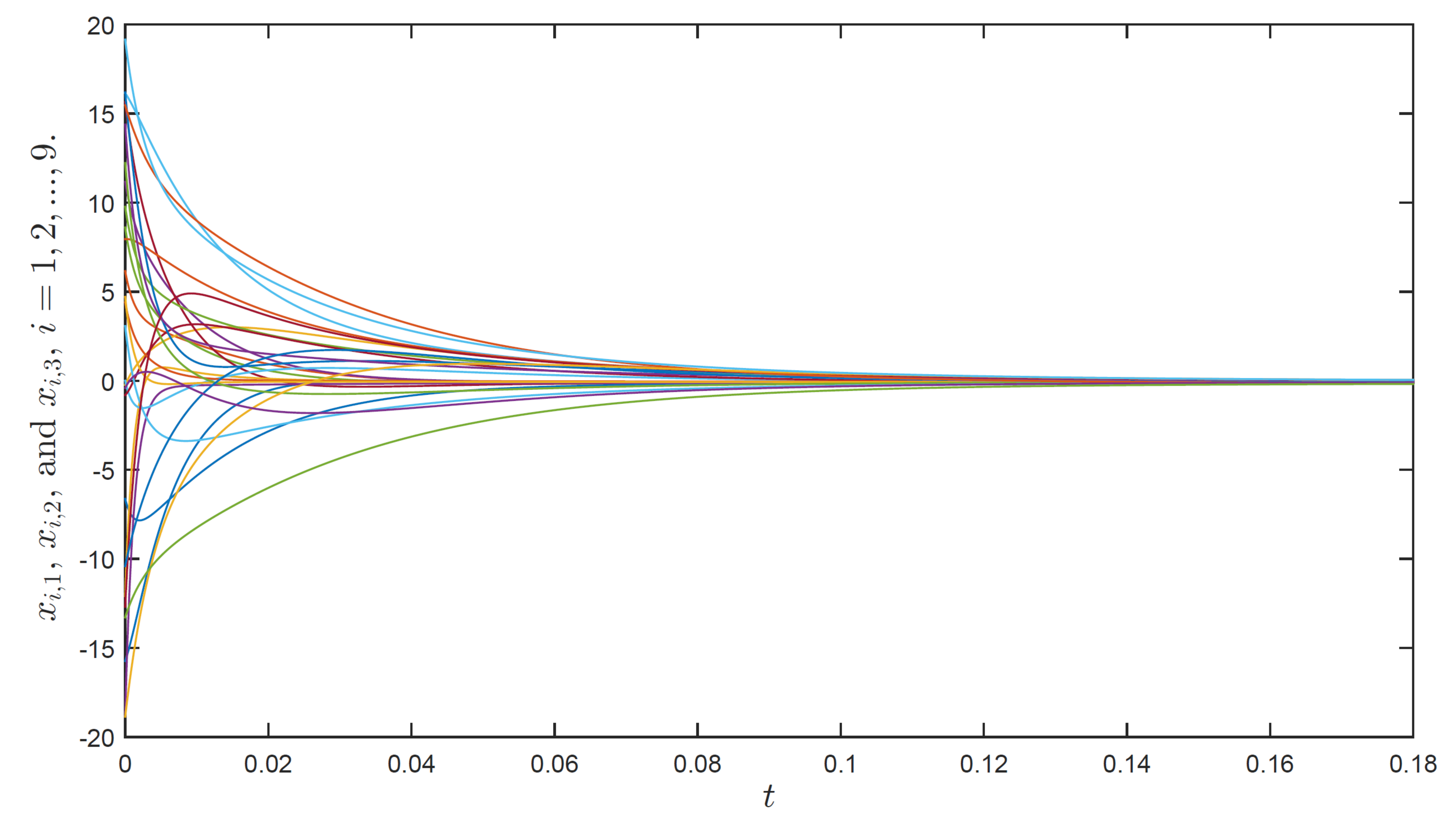}
\end{minipage}%
}

\centering
\subfigure[The node state evolution in the complex network under pinning control scheme $\bm \beta_2$.]
{
\begin{minipage}[c]{0.5\textwidth}
\centering
\includegraphics[width=8.4cm]{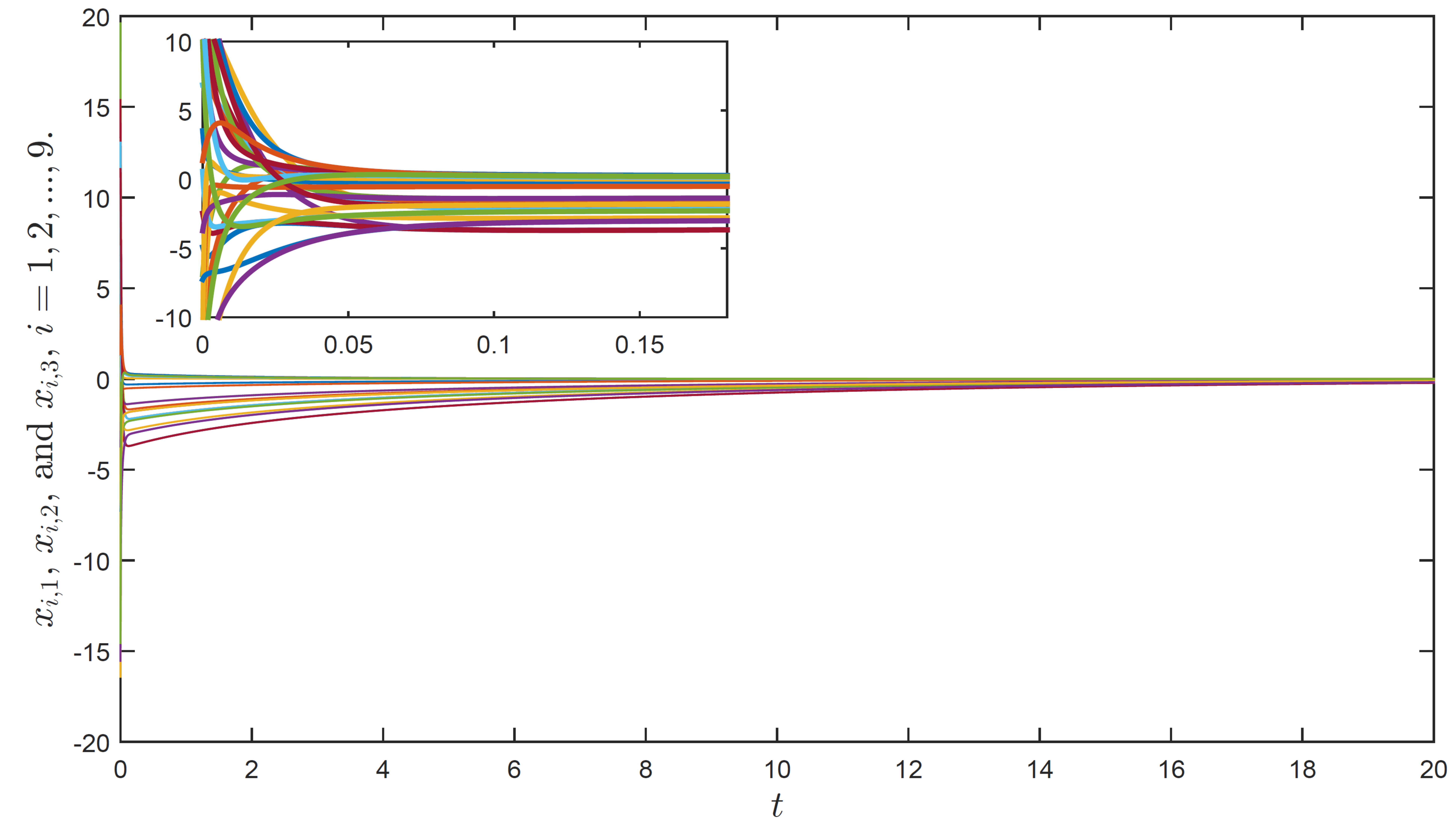}
\end{minipage}%
}

\caption{Comparison of the node state evolutions in the complex network under two  pinning control schemes.} \label{fig:simu1}

\end{figure}

We first consider the pinning control problem in the case of free selection of nodes as described in Section~\ref{sec:CaseI}. Assume that $v_i=0.1k_i$, which represents that the cost coefficient for controlling node with higher number of connections is larger, and $\mathcal V_{\text{c}}=\{4,6\}$. Based on Theorem~\ref{thm:convex1}, the optimal pinning nodes and the corresponding control gains in terms of minimal total costs at $\bm {\bar x}=[0,0,0]'$ are given by:
\begin{equation*}
  \bm \beta_1=[0,0,0,1.2411,0,1.9855,0,0,0]'.
\end{equation*}
As a comparison, when the cost coefficient for node $4$ is large enough, e.g., $v_4>10$, the solution becomes:
\begin{equation*}
  \bm \beta_2=[0,0,0,0,0,11.4341,0,0,0]',
\end{equation*}
i.e., the optimal solution in terms of minimal total costs is only pinning control node $6$ with a lager control gain. A comparison of the node state evolutions in the complex network under two  pinning control schemes is provided in Figure~\ref{fig:simu1}, where both pinning control schemes drive the network to a homogenous state at $\bm {\bar x}=[0,0,0]'$, but $\bm \beta_1$ leads to a faster convergence rate.

Now we consider the case with the attacker. Consider the case where an identical control
gain is adopted. Assume that $\mathcal V_{\text{pin}}=\{4,6,7\}$, $\eta=2$, $\bar c=c$, and:
\begin{equation*}
  \bm {\kappa}\triangleq\big[1,1,1,10,5,6,5,1,1\big]'.
\end{equation*}

\begin{figure}[ht]

\centering

\subfigure[The node state evolution in the complex network without attacks.]
{
\begin{minipage}[c]{0.5\textwidth}
\centering
\includegraphics[width=8.5cm]{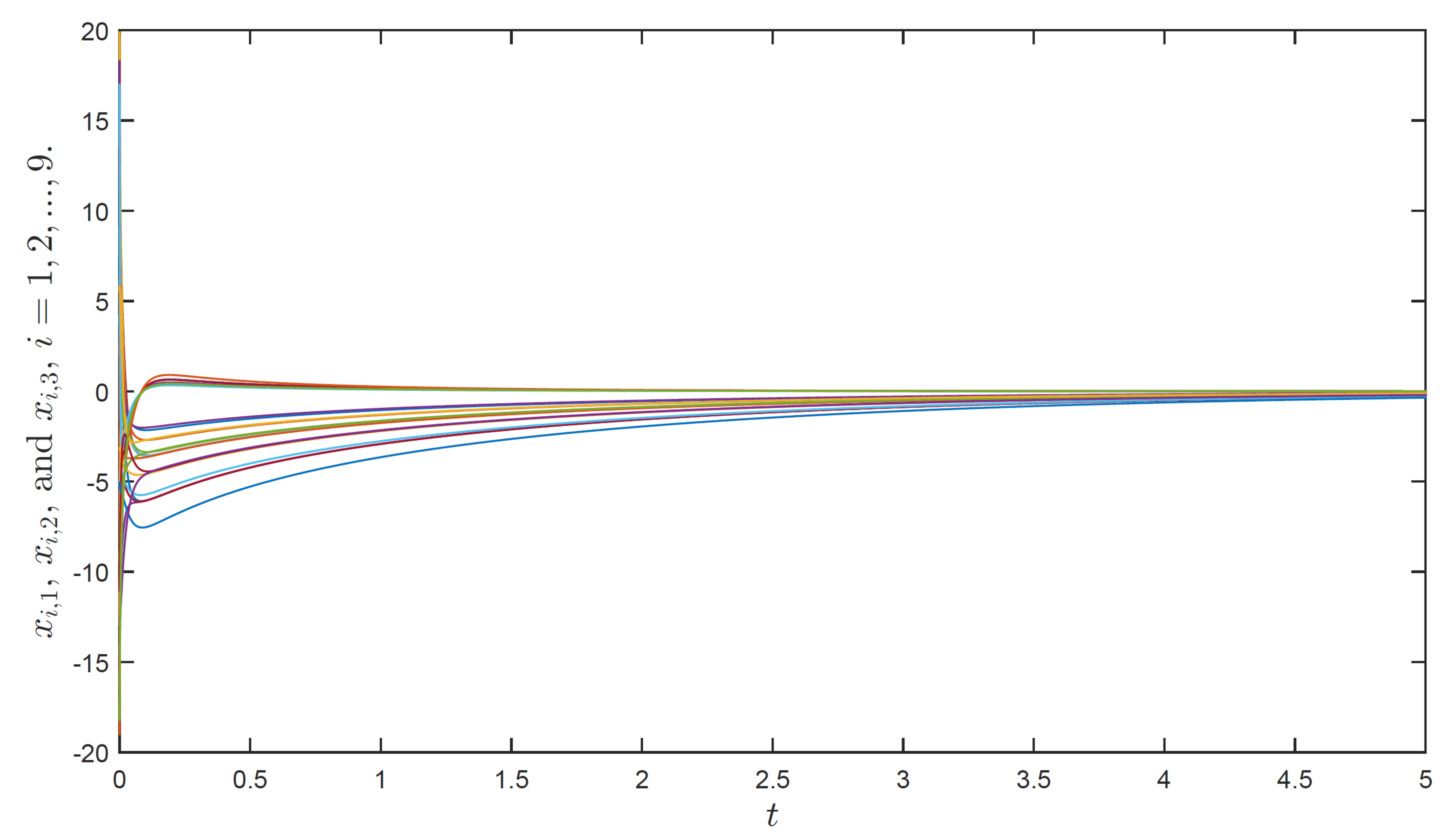}
\end{minipage}%
}

\centering

\subfigure[The node state evolution in the complex network under attacks.]
{
\begin{minipage}[c]{0.5\textwidth}
\centering
\includegraphics[width=8.5cm]{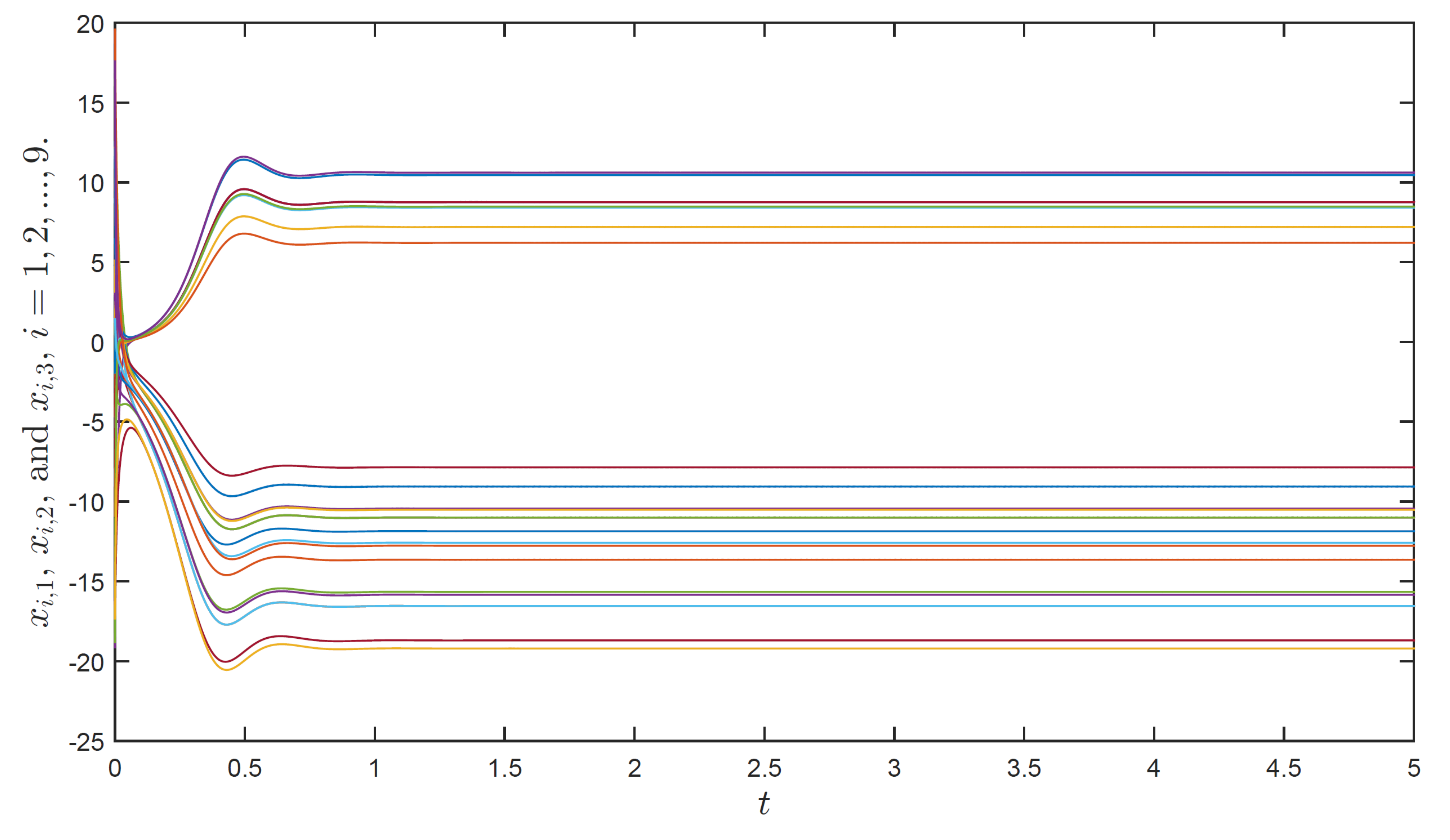}
\end{minipage}%
}

\caption{Comparison of the node state evolutions in the complex network before and after the attacker compromising node $7$.} \label{fig:simu2}

\end{figure}

Based on Theorem~\ref{thm:Solution_stackelberg}, the optimal resource allocation for the defender is given by:
\begin{equation*}
   \bm  \pi_d^\star=[0,0,0,1.5003,0,2.5007,3.0009,0,0]'
\end{equation*}
while the optimal pinning attacking node selection for the attacker $\bm  {\beta}_{\text{attack}}^\star$ is given by:
\begin{equation*}
 \bm  {\beta}_{\text{attack}}^\star=[0,0,0,0,0,0,1,0,0]',
 \end{equation*}
i.e, the attacker can achieve its pinning attacking goal with minimal cost by compromising node $7$. The cost for the attacker is given by $\mathcal U_a=5$ (the cases for fixed budget constraints can be calculated in a similar way). The evolutions of the complex network before and after the attacker compromising node $7$ are shown in Figure~\ref{fig:simu2}.

\section{Conclusion}

In this paper,  a pinning node selection and control gain co-design problem for complex networks was first studied. A necessary and sufficient
condition for synchronization of the pinning controlled network at a homogeneous  state was derived. Based on a quantitative model to describe the pinning costs, we formulated the  pinning node selection and control gain design problem in different scenarios as corresponding optimization problems. Algorithms to solve these problems efficiently were proposed. Based on the developed results, a resource allocation model for the complex network defender and a malicious attacker was described. A Stackelberg game framework was set up to study the behaviour of both sides and the solution to this security game was obtained. Numerical examples and simulations were demonstrated to illustrate the main results.

\bibliographystyle{IEEETran}
\bibliography{D:/Dropbox/Research/0-going/reference}

\end{document}